\documentclass[a4paper,twoside]{article}

\usepackage{epsfig}
\usepackage{subcaption}
\usepackage{calc}
\usepackage{amssymb}
\usepackage{amstext}
\usepackage{amsmath}
\usepackage{amsthm}
\usepackage{multicol}
\usepackage{pslatex}
\usepackage{apalike}
\usepackage[bottom]{footmisc}
\usepackage[hyphens]{url}
\usepackage{hyperref}
\usepackage{SCITEPRESS}     
\usepackage{todonotes}

\begin{document}

\title{Systematically Searching for Identity-Related Information in the Internet with OSINT Tools}

\author{\authorname{Marcus Walkow\sup{1} and Daniela Pöhn\sup{1}}
\affiliation{\sup{1}Universität der Bundeswehr München, Neubiberg, Germany}
\email{\{firstname.familyname\}@unibw.de}
}


\abstract{The increase of Internet services has not only created several digital identities but also more information available about the persons behind them. The data can be collected and used for attacks on digital identities as well as on identity management systems, which manage digital identities. In order to identify possible attack vectors and take countermeasures at an early stage, it is important for individuals and organizations to systematically search for and analyze the data. This paper proposes a classification of data and open-source intelligence (OSINT) tools related to identities. This classification helps to systematically search for data. In the next step, the data can be analyzed and countermeasures can be taken. Last but not least, an OSINT framework approach applying this classification for searching and analyzing data is presented and discussed.
}

\onecolumn \maketitle \normalsize \setcounter{footnote}{0} \vfill

\section{\uppercase{Introduction}}
\label{sec:introduction}

The software company LastPass examined the password behavior of individuals~\cite{blog_psychologie_passwoerter}. According to them, 92 percent know that it is risky to use passwords more than once. Nevertheless, 65 percent always or mostly still use the same password or variations. While financial accounts primarily receive stronger passwords (68 percent), work-related accounts and medical records do not (32 resp. 31 percent). For only 8 percent of the participants, a strong password should not be tied to personal information. According to \cite{zhang_security}, it is possible to predict changes to the password. Consequently, searching for personal information on the Internet may lead to a valid new password. This is even more serious as attacks are increasing, leading to further credentials and personal data being compromised~\cite{verizon}. In organizations, not only one but several digital identities are managed in the identity management system. Typically, users have further accounts, such as web services, where information or credentials can be leaked. Hence, one compromised account in the organization can result in a wider attack.

Open-source intelligence (OSINT) can tackle the problem of the personal factor in passwords and fallback mechanisms. The more knowledge is found about the individual user, the greater the probability that the authentication factor can be derived from it. Hence, the results of a systematic search can warn the user before an incident happens. In order to systematically search for data, a classification is required. In addition, a modular open-source framework helps to apply this classification. The contribution of the paper is two-fold: 1) a classification of data related to identities and identity management systems and 2) an open-source OSINT framework approach based on the classification. This can be utilized to identify possible problematic information.

The paper is structured as follows: Section~\ref{sec:relatedwork} provides an overview of the related work. Section~\ref{sec:osintsearch} introduces and structures OSINT search. The classification is applied by an OSINT framework approach in Section~\ref{sec:casestudy}. The approach is then discussed based on a real-world example in Section~\ref{sec:discussion}. Section~\ref{sec:conclusion} concludes the paper and provides an outlook on future work.

\section{\uppercase{Related Work}}
\label{sec:relatedwork}

Several authors describe OSINT in general. For example, \cite{8954668} provide an overview of OSINT with the basic workflows (collection, analysis, knowledge extraction). Additionally, the authors categorize analysis (lexical, semantic, geospatial, social media) and information (personal, organizational, network). \cite{10.1145/3530977} propose a taxonomy for threat intelligence sharing, which is of limited use for our purpose. The Malware Information Sharing Platform \cite{misp} uses, among others, the categories of blog posts, reports, presentations, news, forums, mailing lists, repositories, and other sources. \cite{8887321} further detail the steps of clustering and correlating data, while \cite{9821829} explain OSINT including web applications, passwords, and emails. Like many other approaches, the authors focus on generic threat intelligence. 

Only a few approaches target OSINT for identities. \cite{7487944} present REAPER, a tool for automated mass credential harvesting. Related to that, \cite{8686093,10.1145/3321705.3329818,BermudezVillalva2018} describe the effects of a password leak. The site Have I been Pwned~\cite{havei} applies a similar approach to warn of password leaks. Social media platforms have changed the way people communicate with each other. At the same time, they are an interesting source for further actions. \cite{9138870,9786736} propose a concept to generate individual password lists based on data gathered by OSINT search. \cite{9315266} go a step further by searching the Internet for information about names, mobile numbers, and email addresses. The authors apply different people's search engines focusing on social media. Similarly, \cite{9848337} performs account matching, extracting user metadata to generate a single report. \cite{akhgar2017open} uses geographic, statistical, and other public sources, while \cite{gibson_acquisition_2016} speaks of unstructured and structured data as well as the type of procurement and the origin of the data.

Especially on GitHub, different OSINT tool lists can be found with \cite{cyberdetective} being the most comprehensive. The author lists, e.\,g., the categories maps, geolocation, and transport; social media; text/sound/video analysis; image search and identification; cryptocurrencies; messengers; search engines; datasets; passwords; emails; nicknames; phone numbers; contact and leak search. OSINT Framework by \cite{osintframework} visualizes different OSINT tools by grouping them into 32 categories, e.\,g., username; email address; images/videos/docs; social networks; instant messaging; people search engines; dating; phone numbers; public records; forums/blogs/IRC; archives; digital currency. Similarly, \cite{osint_tools_2020} lists 7,600 tools and services. Two well-known OSINT tools with open-source and commercial variants are SpiderFoot and Maltego. Although both offer modules related to identities, their main target are domains and networks. Maltego Community Edition (CE) only has one person-related machine, searching for email addresses, whereas SpiderFoot Open Source offers more search options. In addition, full functionality is only available in the paid versions. This shows that further work is required to better protect individual users and organizations with relatively low costs.

\section{\uppercase{OSINT Search}}
\label{sec:osintsearch}

In order to search for compromised identities and further information, which could lead to that state, relevant data first needs to be explored. In the next step, vulnerabilities can be fixed and data be removed to reduce the number of successful fraud attempts. Hence, the goals are the stages of identity research and cyber reconnaissance. We classify possible sources in a systematic way. First, we detail identity search in Section~\ref{sec:osintsearch_identity}. As identity management systems require additional inspections, these are explained in Section~\ref{sec:osintsearch_technical}. Next, we describe all-in-one search tools. Last but not least, we show helpers, which aid in the search and visualization (see Section~\ref{sec:osintsearch_ml}). These sources and helpers can be applied for an extensive OSINT search, using all the different information.

\subsection{Identity Search}
\label{sec:osintsearch_identity}

The data requested during registration (e.\,g., usernames, email addresses, phone numbers, and personal information) can be leaked. Other data, such as relationship status and hobbies, can be used for social engineering and are, therefore, particularly interesting for security issues. Even though multi-factor authentication is increasingly applied, it can be circumvented. Therefore, it is important to reduce published data, described next, which can be found in the following sources~\cite{cyberdetective,misp,9821829}:

\begin{description}
\item[Social Media:] Social media intelligence (SOCMINT) is a sub-branch of OSINT and refers to the information collected from social media websites. The data available can be open to the public or private (cannot be accessed without proper permissions). The content comprises posts/comments, replies, multimedia, social interaction, and metadata.
\item[Search Engines:] Main search engines used by users can be repurposed for OSINT. In addition, meta and specialty search engines are available.
\item[Public Media:] News from newspapers, radio stations, etc. are published online. News digest and discovery tools try to combine specific news.
\item[Public Records:] Reliable and legitimate source of information, e.\,g., registration of a person or financial data of a company.
\item[Repositories:] Codes, snippets, documentation, and other information is published at public repositories, such as GitHub.
\item[Archives:] Website history and capture sites take snapshots of websites that will remain online even if the original page changes or disappears.
\item[Leak Pages:] Pastebin and alternative Pastebin-type sites contain leaks. These leaks are then checked by specific leak pages.
\item[Dark and Deep Web:] Another source for leaks is dark and deep web pages, either information in forums or specific web services.
\item[Further Internet Pages:] This includes forums, blogs, academic resources such as publications, cryptocurrencies, and all other Internet pages.
\end{description}

\subsubsection{Registration Data}

\paragraph{Email}

Email addresses are often used as a substitute for self-chosen usernames or phone numbers. They immediately offer the advantage of an address for the confirmation link. Different tools search for email addresses or check whether an email address exists~\cite{9821829,9315266,cyberdetective}. This includes \texttt{Snov.io}, \texttt{Hunter.io}, \texttt{Skrapp.io}, \texttt{Prospect.io}, \texttt{breachchecker.com}, \texttt{spycloud.com}, and \texttt{haveibeensold.app}. In addition, \texttt{haveibeenpwned.com} searches by email address for leaks.

\paragraph{Username}

Especially at the beginning of Web 2.0, self-chosen user names were common for logging in~\cite{9315266,cyberdetective}. Often, users choose a name, or a variation thereof, with which they have a personal connection. Hence, they often reuse it in the same or in a modified form for other registrations. There are two types of online tools: 1) check whether a profile page exists on various social networks, such as \texttt{whatsmyname.app} and 2) create possible usernames based on entered names, for example, \texttt{namecombiner.com}.

\paragraph{Password}

As users tend to reuse passwords, \texttt{haveibeenpwned.com} lists leaks based on email address. The leaks though can be found at paste sites, dark and deep web~\cite{9821829,7487944,8686093,10.1145/3321705.3329818,BermudezVillalva2018,havei}. In addition, default passwords and password crackers are available online.

\paragraph{Phone Number}

Eliminating the ownership factor simply by knowing the phone number requires additional technologies. However, there are hardly any online services that link a mobile phone number with a name or email address. A classic method is the telephone book, which mainly publishes landline numbers. The latter can be used, for example, via an SMS for multi-factor authentication if no mobile phone number is available. For practical attacks, landline number cloning is more complicated than mobile number cloning.
Some online services provide metadata, such as the provider for a specified phone number~\cite{9315266,cyberdetective}. There is the option of querying cell phone numbers that have been found via Google Dork. The phone number can also be read from social media using suitable crawlers or online services. A possible tool for Instagram, for example, is \texttt{istaunch.com}.

\paragraph{Address}

Personal information such as postal (shipping) addresses can often be found in identity management systems of organizations~\cite{cyberdetective}. This information can be collected online after a leak. In addition, telephone books and public administrations provide further sources. For example, addresses are included in criminal and traffic registers and property searches in the US. \texttt{Hitta.se} is a Swedish search engine that offers telephone directories, addresses, and maps. Last but not least, search engines collect information.

\subsubsection{Further Data}

\paragraph{Texts and Relationships}

Information about social relationships (personal and organizational \cite{8954668}) can provide a plausible background story for social engineering attacks or answers to security questions. Depending on the country, different networks dominate the market (e.\,g., Russia VKontakte). In consequence, several tools are specialized~\cite{cyberdetective}. As an example, \texttt{instahunt.co} looks for usernames in Instagram, while crawlers such as Osintgram automate the quest. In contrast, meta-search engines explore different social networks, search engines, archives, and other websites in their forwarded search queries.\texttt{yasni.de}, for example, focuses on German-speaking countries, \texttt{spokeo.com} addresses the US, and \texttt{social-searcher.com} can be used internationally. In order to provide additional background information, further searches, such as about cryptocurrencies (e.\,g., with \texttt{blockcypher.com}) can be applied.

\paragraph{Image, Video, and Sound}

Users upload several pictures and other material to social media and specific pages~\cite{cyberdetective}. Faces, objects, and logos be recognized in a photo using Google Vision~\cite{google_vision} or Microsoft's face recognition API~\cite{microsoft_facesrecognition}. Tools such as \texttt{reversearch.com} can reverse search or analyze the image, e.\,g., with Sherloq. \texttt{Huntel.io} and further tools analyze the geolocation if published by exchangeable image file format (EXIF) data. Political information, maps, etc. help to locate the material.

\subsection{Technical Search}
\label{sec:osintsearch_technical}

Cyber reconnaissance is a technical investigation aiming to provide attackers with as much information about the target as possible. This includes which (identity) software is being used. On the other hand, publicly available data can be browsed. Hence, the following sources can be utilized~\cite{cyberdetective,misp,9821829}:

\begin{description}
\item[Social Media:] SOCMINT refers to the information (text, multimedia, interaction, metadata) collected from social media websites.
\item[Search Engines:] Main, meta, and specialty search engines search for selected topics.
\item[Public Media:] Online news from original sources.
\item[Public Records:] Reliable and legitimate source of information, e.\,g., financial data of a company.
\item[Repositories:] Data published at public repositories, such as GitHub.
\item[Archives:] Snapshots of sites taken by archives.
\item[Leak Pages:] Information about leaks.
\item[Dark/Deep Web:] Information below the surface.
\item[Further Internet Pages:] All other Internet pages.
\item[Organisation Website:] Organisations provide information online via their organization websites, such as email addresses, roles, and persons.
\item[Network:] The organization's network and servers offer information about insecure systems, used operating system and software versions, and internet protocol (IP) addresses.
\end{description}

\subsubsection{Unsecured Data}

Unintentionally leaked information, such as application programming interface (API) keys, credentials, and internal information, can be used during the attack lifecycle. With self-built scrappers, crawlers or uniform resource locator (URL) fuzzers such as \cite{kraulhorizon}, and Google Dorks, publicly accessible folders, files, and data are displayed.

\subsubsection{Network Data}

\paragraph{Network Scanner}
In order for the identity management systems to be screened using OSINT, the associated hardware must be found on the Internet. \texttt{Shodan.io} systemically asks for relevant ports and publishes the results in a queryable format~\cite{9660818}. \texttt{Censys.io} provides a similar service. Network scanners such as the Network Mapper (NMAP) can be used to find out more about the IT infrastructure. 

\paragraph{Application Testing Software}
Tools such as Burp Suite examine the website of the identity management system. The Burp Suite extension security assertion markup language (SAML) Raider focuses on the federated identity management protocol SAML. The Open Authorization (OAuth) scanner extension detects misconfigurations in the protocol implementations of OpenID Connect and OAuth. 

\paragraph{Security Scanner}
If the identity management system software is known, databases such as \texttt{exploit-db.com} display known vulnerabilities and exploits~\cite{cyberdetective}. So-called security scanners such as the Open Vulnerability Assessment Scanner (OpenVAS) thoroughly test the server behind it for possible vulnerabilities. \texttt{pentest-tools.com} provides a collection of such security scanners.

\subsection{All-in-One Search Tools}
\label{sec:osintsearch_allinone}

All-in-one search tools reuse the tools listed above and combine the results across group boundaries~\cite{cyberdetective}. Recon-ng, SpiderFoot, TiDOS, The Harvester, and Maltego are comprehensive representatives. In the case of Maltego and SpiderFoot, the range of tools differs depending on the version. APIs for paid services such as Social Links CE can be integrated into Maltego CE and SpiderFoot HX. In the full version, external services such as \texttt{pipl.com} or People Data Labs are purchasable. Although these tools provide all-purpose searches, such as social media, search engines, dark web, and leak pages, their main focus is on organization networks.

\subsection{Helpers}
\label{sec:osintsearch_ml}

Due to the huge amount of data that can be found on the Internet, advanced techniques are needed to analyze the data and make a pre-selection.

\subsubsection{Machine Learning}

Machine learning is suitable for this task. Thereby, the collected photos can be evaluated by various social media and provide new insights into identities that were not yet obvious through research. Valuable information is also found in (short) messages and other texts on the Internet, where machine learning algorithms help to extract keywords and analyze the context. Microsoft's Text Analytics~\cite{microsoft_textanalyse}, IBM's Watson API~\cite{IBM_watson}, and Google's Natural Language API~\cite{google_nlp} provide such analysis services.

\subsubsection{Natural Language Processing}

The aim of NLP~\cite{noubours2013nlp} is to process natural language and thereby be able to grasp the meaning of texts and language. Just like people, a computer should have eyes and ears to pick up speech and analyze it with the brain, convert it into code or text, and then process it. In NLP, the problem is best addressed through deep learning models, where sufficient learning material is available due to large data collections. For the purpose of the paper, named-entity recognition (NER)~\cite{yang2012mining,8999622}, sentiment analyzes~\cite{10.1145/3383902.3383904}, and text generation~\cite{10.1145/3491102.3502030} are of particular interest. A current text generator is a generative pre-trained transformer (GPT)-3 by OpenAI.

\section{\uppercase{Example and Case Study of an OSINT Framework}}
\label{sec:casestudy}

This section describes our open-source OSINT framework to search for identity-related information (see Section~\ref{sec:osintsearch_identity}). In addition, the technical search detailed in Section~\ref{sec:osintsearch_technical} can be used if the target is an organization. The framework exerts the workflow described by~\cite{8954668}: Data collection (see Section~\ref{sec:casestudy_collection}), data analysis (see Section~\ref{sec:casestudy_analysis}), and knowledge extraction (see Section~\ref{sec:casestudy_extraction}).

\subsection{Overview}
\label{sec:casestudy_overview}

Our OSINT framework has a graphical user interface (GUI) for interaction with the user. Thereby, the user can select different modules for their search. The modules are implemented or attached in the backend, which interacts with the storage (using a pre-defined folder structure) and database. The search results are then displayed in graphs. In order to realize the interactions, the framework Dash was chosen.

\begin{figure}[!htpb]
    \centering
    \includegraphics[width=0.98\linewidth]{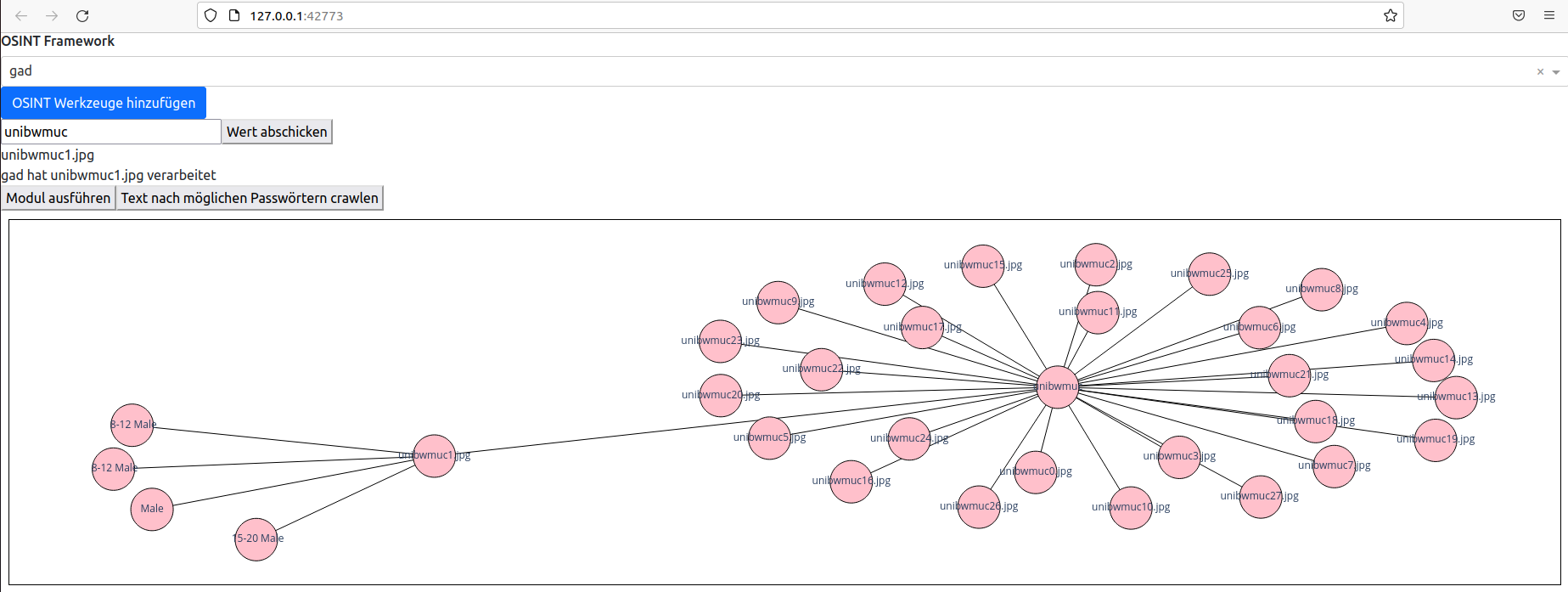}
    \caption{OSINT framework}
    \label{fig:framework}
\end{figure}

Figure~\ref{fig:framework} provides a brief overview of the GUI. In the top line, new values (e.\,g., names, identities, email addresses) are added. This can be combined with modules, which come next. In the big frame below, the results are displayed. In the example within the figure, a node with an image was selected and passed to a module. This evolved into new nodes with further information. In the next section, Figure~\ref{fig:Olaf_Scholz} details the overview on an example. Thereby, the workflow can be iterated.

\subsection{Data Collection}
\label{sec:casestudy_collection}

For the data collection, we use our own and external tools based on Google Dork, scrappers, and crawlers for various sources including social media sites. For example, if a name is entered, possible email addresses are generated and then checked for existence. Next, those valid email addresses are then used to search for social media accounts with Sherlock~\cite{sherlock}. Based on the results, different crawlers download texts, images, and videos as well as further data.
The raw data is either written directly to the project database or stored in the respective folders. In the next step, images and texts are included to be analyzed with suitable tools. Thereby, the phase data collection especially focuses on usernames, texts, and relationships, as well as media. If addresses and phone numbers are part of the social media profiles, then these are also collected.

\subsection{Data Analysis}
\label{sec:casestudy_analysis}

The data analysis again uses external and implemented tools. For example, images are analyzed for geospatial data in EXIF format. Using an API, Google Vision is supposed to recognize texts or faces in images. Images can be further analyzed for location information, such as buildings. The text analysis utilizes an API to Microsoft Text Analysis and the NER extraction tool for the German language. A list of found tokens is returned via the API. In order to receive full words, the words associated with the token are searched in the original text.
Results from the analysis are transferred to the database for knowledge extraction. This phase focuses on texts and media, although the selected words are used as input to generate possible passwords. An attacker could apply these passwords for brute-force attacks. In a defensive scenario, the results may help to rise awareness and improve current passwords.

\subsection{Knowledge Extraction}
\label{sec:casestudy_extraction}

For the extraction of knowledge from the images and texts, the APIs of machine learning algorithms provided by Microsoft and Google are applied. Thereby, emotions among others can be discovered. In order to receive age and sex/gender, two convolutional architectures for fast feature embedding (CAFFE) models with an OpenCV library are used. Here, the pictures are transformed into binary large objects (BLOBs) and transferred to the deep neural network (DNN) of the CAFFE model.
All liable results serve as input for, e.\,g., GPT-NeoX to generate text messages, which could be used by attackers. In addition, possible passwords are created by a custom wordlist generator. This shows, that knowledge extraction requires the described helpers in Section~\ref{sec:osintsearch_ml}.

\section{\uppercase{Discussion}}
\label{sec:discussion}

We discuss our OSINT framework based on an exemplary search, the intended usage, and a brief comparison with other OSINT tools.

\subsection{Applying the OSINT Framework}

To explain how the OSINT framework works, an exemplary search was conducted on German Chancellor Olaf Scholz. This was limited to gender age detection (GAD) of images from his Instagram page and NER of his tweets. After defining the target person of the search, the new node "olafscholz" as the username for Instagram was entered. This is possible as a self-built Instagram crawler was added via the corresponding menu. The framework was informed that the crawler needs information in string type as input, saves data as result, and inserts the file names as nodes. With this self-built crawler, all images were downloaded.

\begin{figure*}[!htpb]
    \centering
    \includegraphics[width=0.9\linewidth]{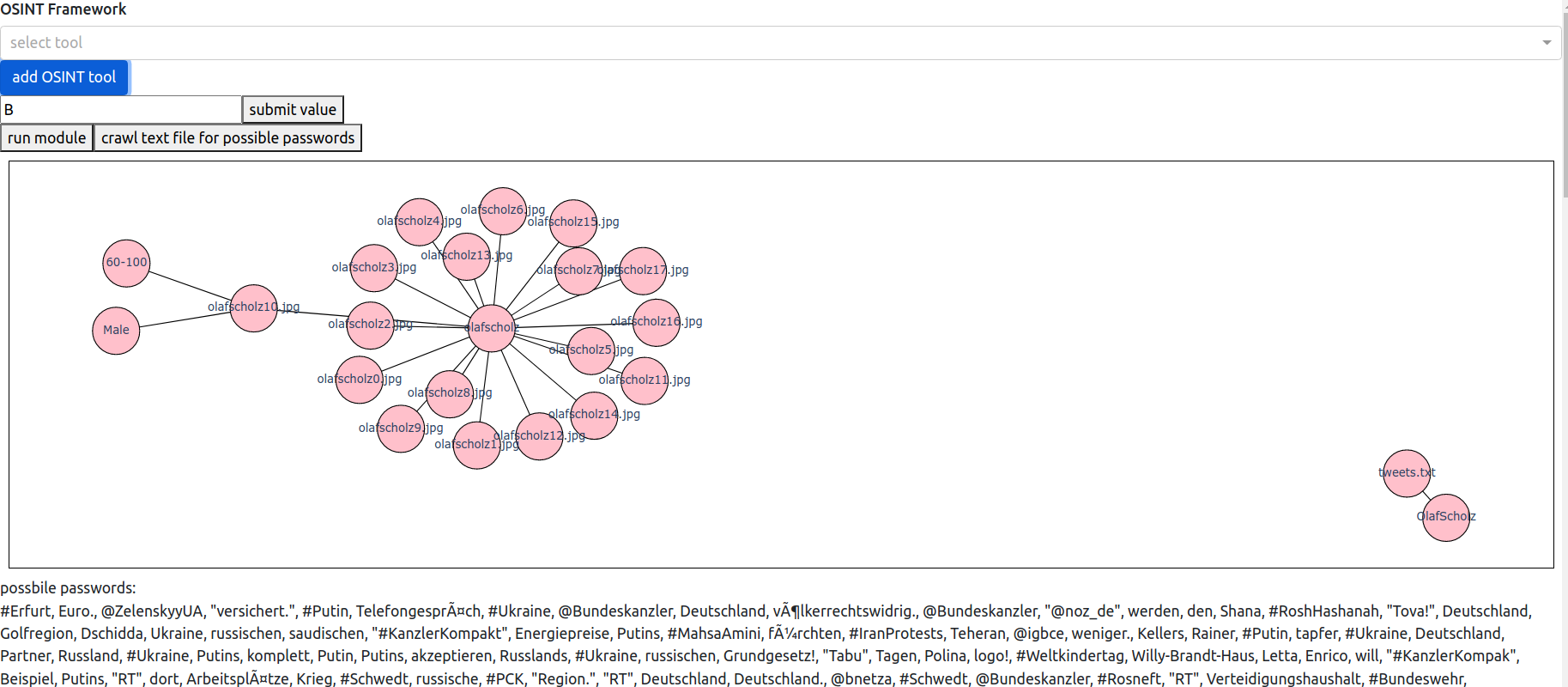}
    \caption{Research on Olaf Scholz}
    \label{fig:Olaf_Scholz}
\end{figure*}

In Figure~\ref{fig:Olaf_Scholz}, 19 found images are displayed. Depending on the results, this overview can get too crowded. In the future, this framework will provide better placement and formatting of the nodes and edges; this may include a reduction of results by grouping.

\subsection{Data Analysis}

For the data analysis, a photo of the results (\texttt{olafscholz10.jpg}) was selected via its representative node and examined with the ML application GAD. Unlike the crawler, GAD requires images as input and returns information. The results of \texttt{olafscholz10.jpg} are inserted as two new nodes on the graph. For Twitter, the tool \texttt{vicinitas.io} is used. The tweets are stored as text files in the "Files" folder and a node is added to the graph. 

By selecting the node and the German NER ML analysis tool, after pressing the 'crawl text file for possible passwords' button, all tweets are examined for named entities. These appear below the graph and are stored in a text file. In later versions, the results will be sorted by frequency and by sentiment analysis according to emotional significance. The text file can be used by programs such as Hydra to reduce the time for a brute-force attack. The assumption behind this is that users choose their passwords with a personal reference to remember them better. However, it should be noted that further cleaning of the words must take place to remove the \# and \@ characters that are typical for Twitter. Furthermore, the German NER also recognizes words as named entities that are none. In the future, information from nodes will be added to the password list, in order to include, for example, Olaf Scholz's wife Britta Ernst as well as other information found about her.

The automated analysis of photos that supports research has been demonstrated with GAD. This capability becomes more helpful when extended with other services such as Google Vision. In the next step, we plan to test the framework on more private individuals as they are typically not aware of the external effects on their posts.

\subsection{Comparison and Limitations}

The basic functioning of collecting and gaining information about identities has been explained. On the technical side, however, there are still some limits and obstacles in comparison to the established tools. The search can only search at locations, where modules with APIs are already written. Further APIs still have to be integrated. In case of APIs are not possible, the search becomes cumbersome. This is also one limitation of existing tools. In comparison, Maltego CE found four unrelated email addresses, whereas SpiderFoot Open Source said the person exists. The latter result did not change when including a Google API.

As programs return different data types, a workaround for Python-based programs was created. From the \texttt{subprocess.run()} Python method used, each output of the executed program is recorded as a string. For outputs in list format, an interpreter was written that turns the string back into a list. Further interpreters for other Python typical data formats are planned. However, here lies a possible weakness that affects user-friendliness. If developers were to use proprietary data types for the output, users would have to write their interpreter or information extraction process. A first idea would be to create a menu, as envisaged for the integration of APIs, in which users insert a tool output and mark which information is relevant. The framework should then recognize this, save it, and build an interpreter.

\section{\uppercase{Conclusion and Outlook}}
\label{sec:conclusion}

OSINT unearths information just waiting to be discovered - either by an individual/organization or an attacker. In order to master the flood of information, classification is necessary for a systematic search. This paper provides a systematic classification for identity and technical data, which is based on a literature review and available tools. In addition, all-in-one search tools and helpers are described. The classification is applied by the open-source OSINT framework approach, which covers the phases of data collection, data analysis, and knowledge extraction. The OSINT framework approach is then discussed based on a targeted search on Olaf Scholz. It shows that open-source tools are possible, though require additional work to produce similar or better results than established tools with a focus on networks.

In order to provide a comprehensive tool for identity protection, further sources will be added in future work. In addition, we plan a user study on the usability and success rate, comparing the results with other open-source and commercial tools. At the same time, countermeasures to hide one's information will be outlined. This OSINT framework will then be extended for organizational purposes.

\bibliographystyle{apalike}
{\small
\bibliography{osint}}

\end{document}